\documentclass[runningheads]{llncs} 

\usepackage{graphicx}
\usepackage{xcolor}
\usepackage{hyperref}
\usepackage{amsthm}
\usepackage{amsmath}

\usepackage{orcidlink}
\usepackage{tablefootnote}
\usepackage[sort,nocompress]{cite}

\usepackage{threeparttable}
\usepackage{makecell}

\newlength{\oldparindent}
\newcommand{\myindent}{\hspace{\oldparindent}}

\theoremstyle{definition}
\newtheorem{owndefinition}{Definition}

\begin{document}

\title{Business Document Information Extraction: Towards Practical Benchmarks}

\author{
    Matyáš Skalický\,\orcidlink{0000-0002-0197-7134}
    \and
    Štěpán Šimsa\,\orcidlink{0000-0001-6687-1210}
    \and
    Michal Uřičář\,\orcidlink{0000-0002-2606-4470}
    \and
    Milan Šulc\,\orcidlink{0000-0002-6321-0131}
}
\authorrunning{M. Skalický, Š. Šimsa, M. Uřičář, M. Šulc}

\institute{Rossum.ai\\\email{\{matyas.skalicky, stepan.simsa, michal.uricar, milan.sulc\}@rossum.ai}}

% typeset the header of the contribution
\maketitle

% The abstract should briefly summarize the contents of the paper in 15--250 words.
\begin{abstract}
Information extraction from semi-structured documents is crucial for frictionless business-to-business (B2B) communication.
While machine learning problems related to \textit{Document Information Extraction} (IE) have been studied for decades, many common problem definitions and benchmarks do not reflect domain-specific aspects and practical needs for automating B2B document communication. We review the landscape of Document IE problems, datasets and benchmarks. We highlight the practical aspects missing in the common definitions and define the \textit{Key Information Localization and Extraction} (KILE) and \textit{Line Item Recognition} (LIR) problems. There is a lack of relevant datasets and benchmarks for Document IE on semi-structured business documents as their content is typically legally protected or sensitive. We discuss potential sources of available documents including synthetic data.
\end{abstract}

\section{Introduction}
\vspace{-1.0mm}
The majority of B2B communication takes place through the exchange of  \textit{semi-structured}\footnote{The term \textit{semi-structured documents} is commonly used in different meanings: Some use it for text files containing semi-structured data~\cite{yi2000classifier}, such as XML files. We use the term to refer to visually rich documents without a fixed layout \cite{riba2019table}.} 
\textit{business documents} such as invoices, purchase orders and delivery notes. Automating information extraction from such documents has a considerable potential to reduce repetitive manual work and to streamline business communication. There have been efforts to provide standards for electronic data interchange of business document metadata~\cite{berge1994edifact,meadows2004universal,bosak2006universal}. Despite, e.g., electronic invoices taking place rapidly~\cite{cristani2018future}, the standards did not get globally adopted, none of them prevails, and most are not inter-operable~\cite{EUdirective201455}.

Machine learning (ML), natural language processing (NLP), and computer vision problems related to \textit{Document Understanding} and \textit{Document IE} have been studied for decades. Despite the major potential of IE from semi-structured business documents, published research on \textit{Document IE} often focuses on other domains \cite{zhong2019publaynet,stanislawek2021kleister,ChenZCXWW20,ChoAHP18,kardas2020axcell,DBLP:conf/icdar/JobinMJ19}, and many of the defined tasks and benchmarks do not reflect domain-specific ML aspects and pitfalls of IE from semi-structured business documents. Publications dealing with business documents typically use private datasets \cite{katti2018chargrid,denk2019bertgrid,holt2018extracting,palm2019attend,palm2017cloudscan,liu2016unstructured,DBLP:journals/corr/abs-1903-12363,schuster2013intellix}, hindering reproducibility and cross-evaluation. This is caused by the absence of a large public dataset of semi-structured business documents, noted by several authors~\cite{palm2017cloudscan,sunder2019one,dhakal2019one,krieger2021information}.

The contributions of this position paper are threefold: first, we provide a review of IE problems, datasets and benchmarks relevant to semi-structured business documents. Second, we identify unaddressed aspects of the tasks and formulate new definitions for \textit{Key Information Localization and Extraction} and \textit{Line Item Recognition}. Third, we stress the lack of a large-scale dataset of semi-structured business documents and we discuss potential sources of documents for such dataset.

\vspace{-1mm}
\section{Document Information Extraction Problems} \label{sec:problems}
\vspace{-.5mm}
\subsection{Key Information Extraction (KIE) and Localization (KILE)} \label{sec:problems_kie}

Most formulations of KIE come from NLP, where it is usually defined as the extraction of values of a fixed set of entities/classes from an unstructured text source into a structured form~\cite{LampleBSKD16,YuLQG020,li.30.1.03nad,huang2019icdar2019,YuLQG020}. Based on the document representation, Garncarek et al.~\cite{GarncarekPSTHTG21} categorize KIE into three groups: (i) sequence-based (working with serialized text~\cite{jiang2012information}), (ii) graph-based (modeling each doc./page as a graph with nodes corresponding to textual segments and their features~\cite{d2018field,krieger2021information,holecek2021learning}), and (iii) grid-based (treating documents as a 2D grid of token embeddings~\cite{katti2018chargrid,denk2019bertgrid}). Sequence-based KIE is closely related to \textit{Named Entity Recognition} (NER)~\cite{LampleBSKD16} --- a sub-task of KIE~\cite{YuLQG020,MajumderPTWZN20} dealing with sequence tagging problems. Borchmann et al.~\cite{borchmann2021due} say that (end-to-end) KIE, unlike NER, does not assume that token-level annotations are available. The task is also referred to as Slot Filling~\cite{palm2017cloudscan}, meaning that a pre-defined slot is filled with the extracted text.

The common definitions of KIE, as well as some of the datasets~\cite{stanislawek2021kleister,borchmann2021due}, do not require the location of the extracted information within the document. While the localization is typically not crucial w.r.t. to the downstream task, it plays a vital role in applications that require human validation. We extend the definition by explicitly including the localization:

\begin{owndefinition}[KILE] \label{def:kile}
Given a document, the goal of \textit{Key Information Localization and Extraction} (KILE) is to localize (e.g., by a bounding box) fields of each pre-defined category (\textit{key}), to read out their values, and to aggregate the values to extract the key information of each category.
\end{owndefinition}
Compared to \textit{Semantic Entity Recognition}, as defined by Xu et al. \cite{abs-2104-08836}, bounding boxes in KILE are not limited to individual words (tokens).
\vspace{-.5mm}
\subsection{Table Extraction and Line Items}

\begin{figure}[tbh]
    \centering
    \vspace{-1mm}
    \includegraphics[width=.70\textwidth, trim=0 32 0 2, clip]{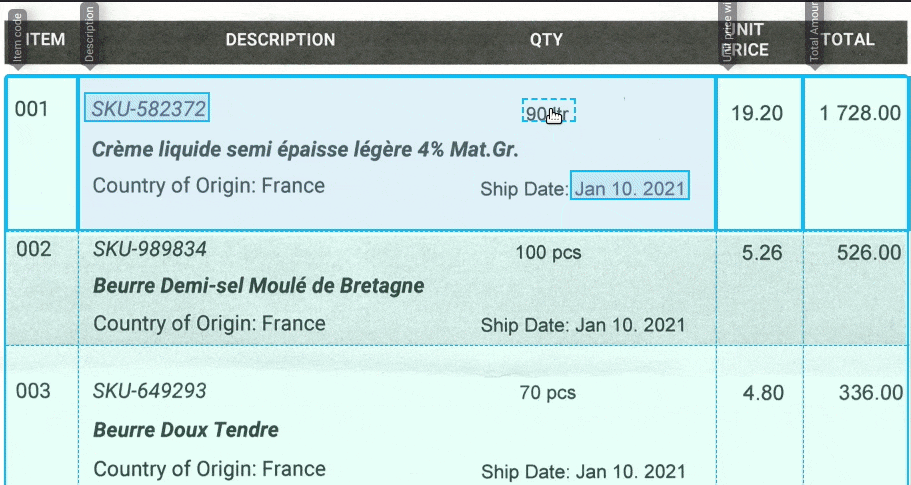}
    \vspace{-1mm}
    \caption{Example of a table structure where field type is not uniquely determined by its column. Source: \href{https://rossum.ai/help/article/extracting-nested-values-line-items/}{https://rossum.ai/help/article/extracting-nested-values-line-items}.}
    %\vspace{-4mm}
    \label{fig:table_structure}
\end{figure}

\textit{Table Understanding} \cite{holevcek2019table} and \textit{Table Extraction} (TE)~\cite{GobelHOO13,ZhengB0ZW21} are problems where the tabular structure is crucial for IE. Unlike KIE, which outputs individual fields independently, TE typically deals with a list of (line) items~\cite{denk2019bertgrid,holevcek2019table,palm2017cloudscan,MajumderPTWZN20,BenschPS21}, each consisting of a tuple of fields (e.g., \textit{goods} and \textit{price}).

In simple tables, columns determine the field type and rows determine which item the value belongs to. The table can therefore be represented as a grid \cite{SchreiberAWDA17,tensmeyer2019deep}. A bottom-up approach \cite{PrasadGKVS20,ZhengB0ZW21} can handle more complex tables as in Fig. \ref{fig:table_structure}, without relying on a row or column detection.
Detected cells or fields can be converted to table structure (determining the line items and columns) in a post-processing step, e.g., spatial clustering \cite{ZhengB0ZW21}. Other works~\cite{ZhongSJ20,LiCHWZL20} tackle the table extraction by directly solving an image-to-markup (e.g., XML or \TeX) problem.

We argue that the problem definition should not rely on the structure but rather reflect the information to be extracted and stored. This is close to the problem of detecting the area belonging to a single line item \cite{denk2019bertgrid}. We define \textit{Line Item} (LI) and the task of \textit{Line Item Recognition} (LIR) as follows: 
\begin{owndefinition}[LI] A \textit{Line Item} is a tuple of fields describing a single object instance to be extracted, e.g., a row in a table.
\vspace{-1mm}
\end{owndefinition}
\vspace{-2mm}
\begin{owndefinition}[LIR]
Given an image of a document page or of a table, the goal of \textit{Line Item Recognition} is to detect all LI present in the section, classify them into a fixed set of classes (e.g., \textit{ordered item}, \textit{discount}, \dots) and for every detected LI, localize and extract key information (as in Definition~\ref{def:kile}) related to it.
\end{owndefinition}
%\vspace{-0.5mm}
\noindent
Note that this definition of LIR allows: (i) detection of several tables with different item types, as well as different item types within a single table; (ii) a single field (e.g., a date) to belong to several line items.
\vspace{-.5mm}
\subsection{One-Shot Learning for Information Extraction}
\label{sec:problems_one-shot}
Layouts of business documents vary greatly, even within a single document type. There are thousands of invoice templates available, and vendors often further adjust them to their needs. Systems without the ability of fast re-training are at risk of degraded performance when faced with a shift in the incoming data distribution \cite{hamza2007case}, such as when presented with previously unseen layouts.

Improving IE with each processed document is known as a one-shot \cite{dhakal2019one}~/~online~\cite{stockerl2015online} template matching, case-based reasoning~\cite{hamza2007case}, or configuration-free IE \cite{schuster2013intellix}. This includes systems that reuse annotations of similar documents~\cite{dhakal2019one,schuster2013intellix} or iteratively build and refine a representation of a document class~\cite{hamza2007case,medvet2011probabilistic,rusinol2013field,d2018field}. Annotations of documents' templates are not part of any public IE dataset of sufficient size.

\subsection{Other Document IE Problems and Tasks}
\label{sec:problems_other}
\myindent \textbf{Optical Character Recognition} (OCR) \cite{smith2007overview}, handwritten OCR \cite{he2017beyond}, scene text recognition \cite{yu2020towards,baek2019character}, including (sub)word or text-line level predictions, are standard problems with reviews and comparisons available \cite{hamad2016detailed,memon2020handwritten,nayef2019icdar2019,islam2017survey}. While highly relevant to the document IE, this paper aims at the ``higher-level`` document IE problems, often assuming text extracted from PDF or OCR is available.

\textbf{Document Layout Analysis} (DLA) is typically posed as an object detection problem: given a document page, find the minimum bounding boxes (or other area representation \cite{antonacopoulos2009realistic,clausner2019icdar2019}) of layout elements such as \textit{Paragraph}, \textit{Heading}, \textit{Table}, \textit{Figure}, or \textit{Caption}. Most DLA datasets \cite{ford2003ground,antonacopoulos2009realistic,zhong2019publaynet,clausner2019icdar2019} contain such layout annotations for scientific~/~technical publications and magazines.

\textbf{Extraction of Key-Value Pairs (KVP)} refers to recognizing pairs of linked data items where the key is used as a unique identifier for the value. This task usually consists of semantic labeling and semantic linking \cite{XuL0HW020,abs-2104-08836}. Contrary to KIE, KVP extraction does not require the set of keys to be fixed. It also assumes that both key and value are present in the document. This may be useful, e.g., to extract data from unknown forms. However, in semi-structured business documents, it is pretty standard that the keys of interests (known in advance) are not explicitly present in the document.

\textbf{Question Answering (QA)}, also known as \textit{Machine Reading Comprehension}, is a common problem in information retrieval and NLP. The goal is to automatically answer questions formulated in natural language. Many NLP tasks can be reformulated as QA~\cite{abs-1806-08730,KumarIOIBGZPS16}. Similar to KIE, QA can be extended to incorporate visual information to Visual Question Answering (VQA) \cite{MathewKJ21}. VQA system may also interpret and extract content from the figures, diagrams, and other non-textual elements.

KIE can be formulated as an instance of VQA. However, we typically know which classes of key information should be extracted, rendering the natural language interface unnecessary.

\vspace{-1mm}
\section{Semi-Structured Business Document Datasets}\label{sec:datasets}
Publications on business document IE are often based on private datasets \cite{katti2018chargrid,denk2019bertgrid,holt2018extracting,palm2019attend,palm2017cloudscan,liu2016unstructured,DBLP:journals/corr/abs-1903-12363,schuster2013intellix}. Due to the documents' sensitive content, authors are typically not allowed to share the experimental data. Large third-party sources like common crawl are publicly available; however, re-publishing such data may pose legal issues. For example, a large common crawl dataset of PDF documents by Xu et al.~\cite{abs-2104-08836} was not published, while pre-training on it was crucial for the proposed method, and the C4 dataset~\cite{raffel2019exploring} is shared in the form of code that extracts it directly from Common Crawl.

Publicly available datasets for KI(L)E from business documents are summarized in Table~\ref{table:datasets}. However, most of them are relatively small and only contain a few annotated field types. The two largest datasets consist entirely of receipts. Table~\ref{table:datasets} does not include datasets without KIE annotations --- RVL-CDIP~\cite{harley2015icdar} (classification), FUNSD~\cite{jaume2019} and XFUND~\cite{abs-2104-08836} (no fieldtypes), NIST~\cite{NIST} (forms identification) and DocVQA~\cite{MathewKJ21} (QA) --- and datasets we were not able to download\footnote{For some only the annotations are available, without the original PDFs/images.} \cite{cesarini2003analysis,rastogi2020information,ZhuLHWZLFC20,BaviskarAK21}. 

Borchmann et al. \cite{borchmann2021due} recently joined and re-formulated several existing document IE datasets to build the DUE benchmark for several document understanding tasks on different document domains. DeepForm \cite{deepform2020} and Kleister Charity \cite{stanislawek2021kleister} are the only subsets of DUE with business document KIE annotations.

\begin{table}[tb]
\centering
\caption{Datasets related to KI(L)E from semi-structured business documents.}
\resizebox{\textwidth}{!}{
\begin{tabular}{lllllllll} 
\hline
name                                              & document type      & docs     & fieldtypes  & source    & multipage & lang. & type  \\ 
\hline
WildReceipt       \cite{sun2021spatial}           & receipts           & $1 740$  & $25$        & photo  & no        & en    & KILE \\
Ghega             \cite{medvet2011probabilistic}  & patents/datasheets & $246$    & $11$/$8$    & scan   & yes       & en    & KILE \\
EPHOIE            \cite{wang2021vies}             & chinese forms      & $1 494$  & $10$        & scan   & no        & zh    & KILE \\
CORD              \cite{park2019cord}             & receipts           & $11 000$ & $42$        & photo  & no        & ind   & KILE \\
DeepForm          \cite{deepform2020}             & invoices, orders   & $1 000$  & $6$         & scan   & yes       & en    & KILE \\
Kleister Charity  \cite{stanislawek2021kleister}  & financial reports  & $2 788$  & $8$         & scan   & yes       & en    & KIE  \\
Kleister NDA      \cite{stanislawek2021kleister}  & NDA documents      & $540$    & $4$         & scan   & yes       & en    & KIE  \\
SROIE             \cite{huang2019icdar2019}       & receipts           & $973$    & $4$         & scan   & no        & en    & KIE  \\
\hline
\end{tabular}
}
\vspace{-2mm}
\label{table:datasets}
\end{table}

While there are many datasets for Table Detection and LIR~\cite{FangTTQL12,zhong2019publaynet,SiegelLPA18,GobelHOO13,DBLP:conf/icdar/GaoYJHT17,clausner2019icdar2019,ShahabSKD10,TableBank,SciTSR,DengRM19,QasimMS19,ZhongSJ20,ZhengB0ZW21,abs-2203-01017}, some of them are not accessible anymore~\cite{GobelHOO13,DBLP:conf/icdar/GaoYJHT17,ShahabSKD10,DengRM19}. We find only FinTabNet~\cite{ZhengB0ZW21} and SynthTabNet~\cite{abs-2203-01017} to be relevant to us by covering complex financial tables. 

\vspace{-0.5mm}
\section{Where to Get More Documents}
\label{sec:more_documents}

\paragraph{Publicly Available Documents.}
Business documents are typically not shared publicly due to their private content, often including confidential and personal information. There are exceptions to this rule --- e.g., institutions such as governments or charities have to make certain financial documents publicly available for transparency reasons. Databases of such documents have already been used to create public datasets for document IE: Several datasets\footnote{\label{footnote:subsets}Some of the datasets are subsets: FUNSD~\cite{jaume2019} $\subset$ RVL-CDIP~\cite{harley2015icdar} $\subset$ IIT-CDIP~\cite{lewis2006building}.} --- IIT-CDIP~\cite{lewis2006building}, RVL-CDIP~\cite{harley2015icdar}, FUNSD~\cite{jaume2019}, and DocVQA~\cite{MathewKJ21} --- were built from documents from the UCSF Industry Documents Library\footnote{A large proportion of the UCSF Industry Document Library are old documents, often written on a typewriter, which presents a domain shift w.r.t. today's documents.}~\cite{IndustryDocumentsLibrary}. Annual Reports of the S\&P 500 companies~\cite{SaP500b} were used to create FinTabNet~\cite{ZhengB0ZW21}. Non-disclosure agreements from the EDGAR\footnote{Automated crawling of the site not allowed: \scriptsize \url{https://www.sec.gov/os/accessing-edgar-data}} database \cite{EDGAR}, collected for the U.S. Securities and Exchange Commission, were used for the  Kleister-NDA~\cite{stanislawek2021kleister} dataset. The DeepForm dataset~\cite{deepform2020} consists of documents related to broadcast stations from the FCC Public Inspection Files \cite{PublicInspectionFiles}. Financial records from the Charity Commission~\cite{CharityCommission} were used to create the Kleister-Charity dataset~\cite{stanislawek2021kleister}. Several QA datasets \cite{ChaudhrySGMBJ20,MethaniGKK20,ZhuLHWZLFC20} were also collected from open data sources~\cite{worldbank,VisualizeDataGovIn,AnnualReports}. Other datasets were build via web search \cite{mathew2022infographicvqa,sun2021spatial,DBLP:journals/corr/abs-2101-09465}, from Common Crawl\footnote{CC-MAIN-2022-05 contains almost 3 billion documents out of which 0.84\% are PDFs~\cite{CommonCrawlMIME} -- however, most of them are not semi-structured business documents.}~\cite{abs-2104-08836,raffel2019exploring}, Wikipedia  \cite{ChenCSWC21,chen2019tabfact,ChenZCXWW20}, or platforms for sharing scientific papers~\cite{kardas2020axcell,zhong2019publaynet,DBLP:conf/icdar/JobinMJ19}.

\vspace{-1mm}
\paragraph{Synthetic documents.}
Manual annotation is expensive, and the collection of data from public sources may be limited by the presence of personal data or intellectual property. This reasoning calls for leveraging synthetic datasets. Xu et al.~\cite{abs-2104-08836} manually replaced the content of publicly available documents with synthetic data. Bensch et al.~\cite{BenschPS21} generate synthetic invoice documents automatically. However, we observe that the generated invoices have a plain style and do not resemble the distribution of visual layouts of real business documents. Nassar et al. \cite{abs-2203-01017} synthesized a table dataset of four appearance styles based on existing datasets \cite{ZhongSJ20,ZhengB0ZW21,LiCHWZL20}.

We consider three ways to define layouts to be filled with synthetic data: (i) manual design of layouts which can be used to generate a high number of documents with different semantically-matching values, but costly at scale with increasing num. of layouts, (ii) extraction from public documents followed by sensitive anonymization like in~\cite{abs-2104-08836}, and (iii) using a generative model, e.g., to generate realistic layouts dissimilar to those already present in the dataset. We consider such a problem statement an interesting open research problem.

% \vspace{-0.5mm}
\section{Discussion and Future Work}

We argue that the problems of KILE and LIR, as defined in Sec.~\ref{sec:problems}, are crucial for automating B2B document communication where key information must be extracted from localized fields and line items. The review of public datasets in Sec.~\ref{sec:datasets} shows that --- except for receipts \cite{sun2021spatial,park2019cord,huang2019icdar2019} --- semi-structured business documents like invoices, orders, and delivery notes are underrepresented in document IE. Based on manual inspection of selected documents from publicly available sources in Sec.~\ref{sec:more_documents}, we noticed the distribution of documents differs significantly among different sources. An ideal dataset should cover a large variety of visual styles and layouts and provide diagnostic subsets \cite{borchmann2021due} to differentiate errors in various special cases. Due to high annotation costs and possibly legally protected content of business documents, synthetic data are a potentially affordable alternative for building a large-scale dataset. While synthetic data have been proven successful for OCR \cite{li2019synthesizing}, the potential of data synthesis for business document IE has not yet been fulfilled: existing attempts either target other tasks and document types \cite{abs-2104-08836} or do not reflect the rich visual distribution of semi-structured business documents \cite{BenschPS21}. An advantage of generating synthetic documents of a given layout is the known layout annotation for benchmarking one-shot information extraction.

To enable benchmarking of information extraction on data and tasks highly relevant to real-world application scenarios, in our future work, we are preparing a large-scale public dataset of semi-structured business documents, following the observations and points made in this paper.

\bibliographystyle{splncs04}
\bibliography{bibliography}
\end{document}